\pdfoutput=1
\documentclass[notitlepage,aps,nofootinbib,
letterpaper,prl,twocolumn,showpacs,amsmath,
amssymb,amsfonts,long,superscriptaddress,preprintnumbers]{revtex4}
\usepackage{amsmath,amssymb,amsthm,amsxtra,overpic,bbm,bm,epsfig,subfigure}
\usepackage{simplewick}
\usepackage{color}
\usepackage{epstopdf}
\usepackage{epsfig}
\usepackage{comment}
\usepackage{graphicx,color,soul}
\usepackage{multirow}
\usepackage{verbatim}
\usepackage{enumitem}
\usepackage[linkcolor=red]{hyperref}
\usepackage{lipsum}

\newcommand{\prlsection}[2]{{\it\textbf{#1}{#2}}---}
\makeatletter
\newcommand*{\balancecolsandclearpage}{%
	\close@column@grid
	\cleardoublepage
	\twocolumngrid
}
\makeatother

\begin{document}

\title{Probing Cosmic Axions through Resonant Emission and Absorption \\ in Atomic Systems with Superradiance}

\author{Guo-yuan Huang}
\email{huanggy@ihep.ac.cn}
\affiliation{Institute of High Energy Physics, Chinese Academy of
Sciences, Beijing 100049, China}
\affiliation{School of Physical Sciences, University of Chinese Academy of Sciences, Beijing 100049, China}

\author{Shun Zhou}
\email{zhoush@ihep.ac.cn}
\affiliation{Institute of High Energy Physics, Chinese Academy of
Sciences, Beijing 100049, China}
\affiliation{School of Physical Sciences, University of Chinese Academy of Sciences, Beijing 100049, China}

\date{\today}

\begin{abstract}
The $\mu$eV-mass axion is one of the most promising candidates for cold dark matter, and remains to be a well-motivated solution to the CP problem of Quantum Chromodynamics (QCD) via the Peccei-Quinn mechanism. In this paper, we propose a novel method to detect the dark-matter axions in our galaxy via the resonant emission $|{\rm e}\rangle \to |{\rm g}\rangle + \gamma + \gamma^{\prime}_{} + a$ (or absorption $a + |{\rm e}\rangle \to |{\rm g}\rangle + \gamma + \gamma^{\prime}_{}$) in an atomic system with superradiance, where $|{\rm e}\rangle$ and $|{\rm g}\rangle$ stand for the excited and ground energy levels of atoms, respectively. A similar process via $|{\rm e}\rangle \to |{\rm g}\rangle + \gamma + a$ (or $a + |{\rm e}\rangle \to |{\rm g}\rangle + \gamma$) is also put forward to probe the axion-electron coupling.
For the nominal experimental setup assuming a background-free environment, most of the parameter space for typical QCD axion models can be covered with parahydrogen molecules or ytterbium atoms. However, the background in a realistic experimental setup remains to be a major issue that needs to be solved in future studies. Searching for better atomic or molecular candidates may be required for a bigger signal-to-noise ratio.
\end{abstract}

\preprint{}

\maketitle

\prlsection{Introduction}{.}%
More than forty years ago, Peccei and Quinn (PQ) proposed an appealing solution to the CP problem of Quantum Chromodynamics (QCD) by introducing a dynamical scalar field and imposing a global ${\rm U}(1)^{}_{\rm PQ}$ symmetry on the whole Lagrangian~\cite{Peccei:1977hh, Peccei:1977ur}. It was Weinberg~\cite{Weinberg:1977ma} and Wilczek~\cite{Wilczek:1977pj} who shortly discovered that a Nambu-Goldstone boson, i.e., the axion, arose from the spontaneous breaking of the PQ symmetry at some high-energy scale. Although the original model with the PQ symmetry spontaneously broken at the electroweak scale $\Lambda^{}_{\rm EW} \equiv 10^2~{\rm GeV}$ has been ruled out, the ``invisible" axion models, such as the KSVZ model~\cite{Kim:1979if, Shifman:1979if} and the DFSZ model~\cite{Zhitnitsky:1980tq, Dine:1981rt}, are still attracting a lot of attention. Apart from providing a solution to the strong CP problem, these models also indicate that axions can be a good candidate for cold dark matter in our Universe~\cite{Preskill:1982cy, Abbott:1982af, Dine:1982ah, Davis:1986xc, Kim:1986ax}, and can be detected in realistic experiments~\cite{Sikivie:1983ip, Cheng:1987gp, Turner:1989vc, Raffelt:1990yz}. An excellent overview of possible experimental methods to probe axions has recently been presented in Ref.~\cite{Irastorza:2018dyq}.

\begin{figure}[t!]
	\begin{center}
		\hspace{-1cm}
		\includegraphics[width=0.525\textwidth]{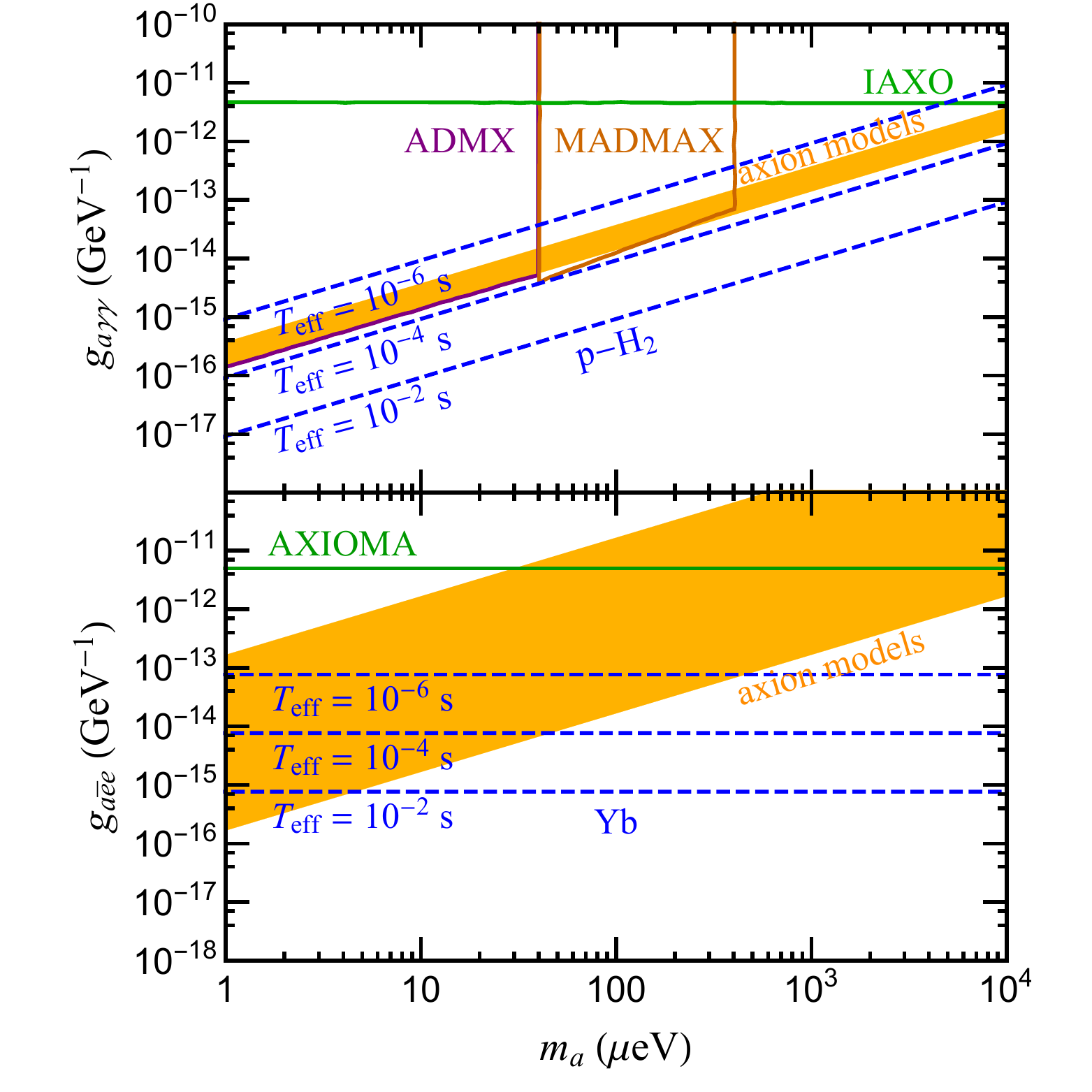}
	\end{center}
	\vspace{-0.3cm}
	\caption{The $3\sigma$ bounds on the axion-photon coupling $g^{}_{a\gamma\gamma}$ (upper panel) and the axion-electron coupling $g^{}_{a \overline{e} e}$ (lower panel) for different axion masses, where a background-free environment has been assumed and $T^{}_{\rm eff}$ denotes the exposure time. See the text for the details of the experimental setup. The sensitivities of projected experiments, such as ADMX \cite{Stern:2016bbw}, MADMAX \cite{TheMADMAXWorkingGroup:2016hpc,Brun:2019lyf}, IAXO \cite{Armengaud:2014gea,Armengaud:2019uso} and AXIOMA \cite{Braggio:2017oyt,Borghesani:2015bla}, are given for comparison.}
	\label{fig:sensitivity}
\end{figure}

Due to the instanton effects, the axion acquires a small mass $m^{}_a$ from the explicit PQ symmetry breaking at low energies. In a generic axion model, the axion mass $m^{}_a$ and decay constant $f^{}_a$ (i.e., the energy scale of spontaneous PQ symmetry breaking) are related to each other via $m^{}_a \cdot f^{}_a \simeq 6.0~{\rm \mu eV} \cdot 10^{12}~{\rm GeV}$. The experimental searches for dark matter axions mainly rely on their couplings to photons and fermions~\cite{Raffelt:2006cw}, i.e.,
\begin{align} \label{eq:Lagrangian}
{\cal L}^{}_a = \frac{1}{4} g^{}_{a\gamma\gamma} F^{}_{\mu \nu} \tilde{F}^{\mu\nu} a + g^{}_{a \overline{f} f} \overline{f} \gamma^{\mu}\gamma^{}_5 f \partial_{\mu}a \; ,
\end{align}
where $F^{}_{\mu \nu}$ denotes the strength tensor of the electromagnetic field, and $\tilde{F}^{}_{\mu \nu}$ its dual. In a specific axion model, the coupling constants $g^{}_{a\gamma\gamma}$ and $g^{}_{a \overline{f} f}$ can be expressed in terms of the PQ symmetry breaking scale $f^{}_a$. For instance, one obtains $g^{}_{a\gamma\gamma} = -\alpha(4+z)/[3\pi f^{}_a (1+z)]$ in the KSVZ model and $g^{}_{a\gamma\gamma} = \alpha z/[\pi f^{}_a (1+z)]$ in the DFSZ model, respectively, where $\alpha$ is the electromagnetic fine-structure constant and $z = m^{}_u/m^{}_d \simeq 0.56$ is the up/down quark mass ratio~\cite{Raffelt:2006cw}. Nevertheless, in other extensions of the standard model, the axion-like particles (ALPs) are predicted, for which the mass-coupling relation is not expected~\cite{Jaeckel:2010ni}. Generally, we take the effective coupling constants $g^{}_{a\gamma\gamma}$ and $g^{}_{a\overline{f}f}$ as free parameters, which are independent of the masses of axions or ALPs. So far, all the experimental searches for axions and ALPs in cosmology, stars and terrestrial laboratories come out with null signals, leading to very restrictive constraints on the masses and couplings (see, e.g., Ref.~\cite{Tanabashi:2018oca}, for a review). It is worthwhile to point out that the QCD axions within a particular mass region $1~{\rm \mu eV} \lesssim m^{}_a \lesssim 10^4~{\rm \mu eV}$ survive all the experimental constraints, and can make up the entire cold dark matter. For this reason, many interesting ideas have been proposed to test the QCD axion models in this region~\cite{Graham:2015ouw,Stern:2016bbw,TheMADMAXWorkingGroup:2016hpc,Brun:2019lyf,Millar:2016cjp,Armengaud:2014gea,Armengaud:2019uso,Sikivie:2014lha,Braggio:2017oyt,Borghesani:2015bla,Hill:2015kva,Du:2018uak,Arza:2019nta,Lawson:2019brd}.

\begin{figure}[t]
	\begin{center}
		\includegraphics[width=0.4\textwidth]{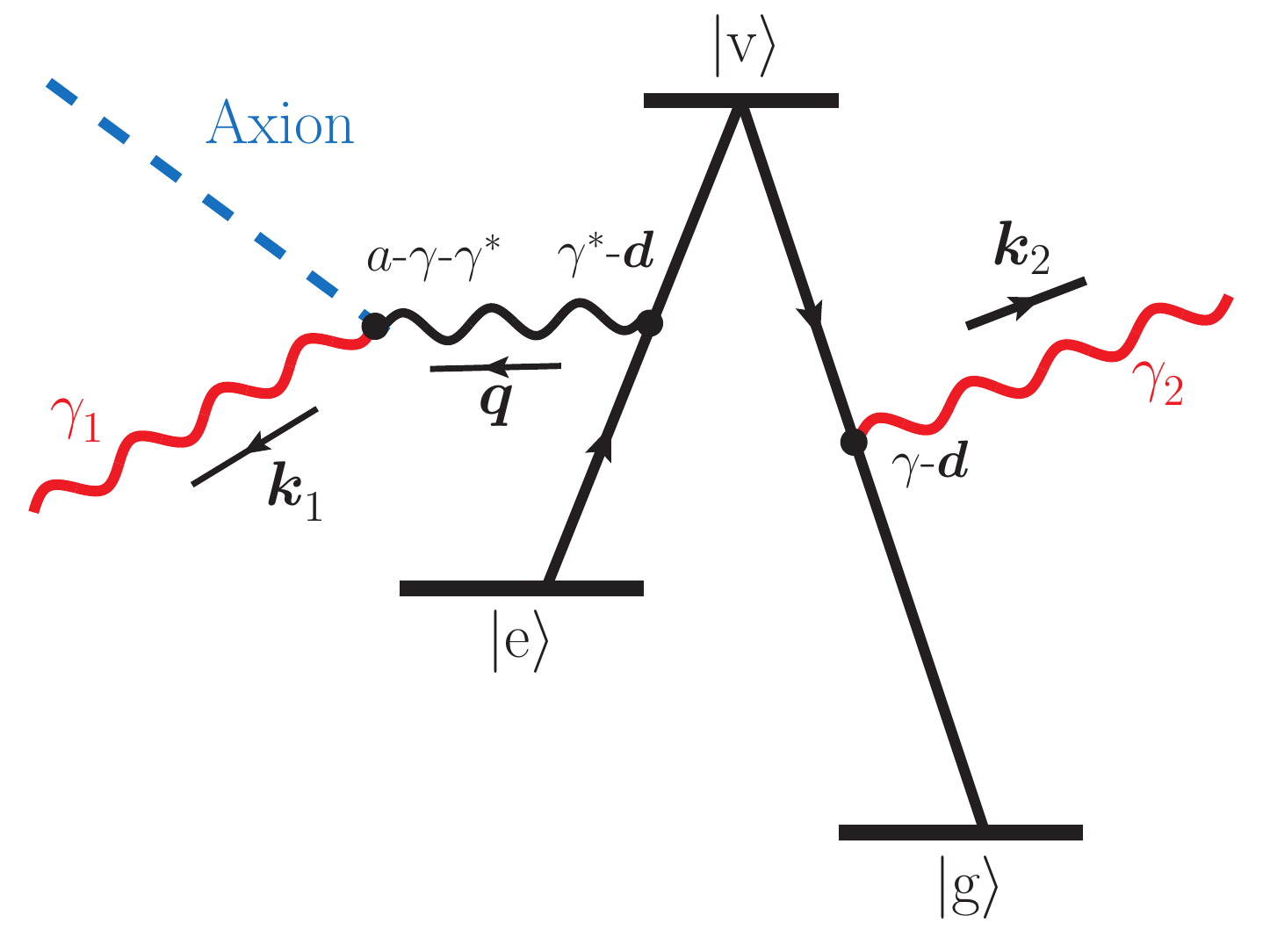}
	\end{center}
	\caption{The atomic transition $|{\rm e}\rangle \to |{\rm g}\rangle + \gamma^{}_{1} + \gamma^{}_{2} + a$ or $a + |{\rm e}\rangle \to |{\rm g}\rangle + \gamma^{}_{1} + \gamma^{}_{2}$ induced by the axion-photon coupling.}
	\label{fig:transition1}
\end{figure}

In this work, we propose to detect cosmic axions via atomic transitions, i.e., triggered emission $|{\rm e}\rangle \to |{\rm g}\rangle + \gamma^{}_{1} + \gamma^{}_{2} + a$ or absorption $a + |{\rm e}\rangle \to |{\rm g}\rangle + \gamma^{}_{1} + \gamma^{}_{2}$ of cosmic axions (TREACA), where $\gamma^{}_{1}$ and $\gamma^{}_{2}$ stand for two outgoing photons and the whole atomic system is stimulated by two beams of triggering lasers. In the assumption that the relevant background is well under control, we examine the experimental sensitivity to the axion-photon coupling for a given axion mass, and the final results are shown in Fig.~\ref{fig:sensitivity}, while the transition process is sketched in Fig.~\ref{fig:transition1}. The main idea is briefly summarized as follows:
\begin{itemize}[noitemsep,topsep=0pt,leftmargin=5.5mm]
\item[(i)] The atomic (or molecular) system of a $\Lambda$-type three energy levels will be considered. The transition between the excited state $\left|\rm e \right>$ and the ground state $\left|\rm g \right>$ can be achieved only via an intermediate state $\left|\rm v \right>$, which is connected to $\left|\rm e \right>$ and $\left|\rm g \right>$ by either electric or magnetic dipole interaction. The whole system of a large amount of such atoms is well prepared in a macroscopically coherent state such that the superradiant emission is realized.

\item[(ii)] The signal for the absorption or emission of the axion with a mass $m^{}_a$ can be identified by observing one of the photons, e.g., $\gamma^{}_{1}$, with an excess or deficiency in frequency $\omega^{}_{1} = E^{}_{\rm eg} -\omega^{}_{2} \pm m^{}_{a}$, where $E^{}_{\rm eg}$ is the energy difference between $\left|\rm e \right>$ and $\left|\rm g \right>$, and $\omega^{}_{2}$ the energy of the other photon. In order to increase the rate, we can use two triggering laser beams corresponding to photon modes of $\gamma^{}_{1}$ and $\gamma^{}_{2}$ to help stimulate the atomic transitions. On this point, our method is different from the one suggested in Ref.~\cite{Yoshimura:2017ghb}, where only one laser trigger is used and the transition rate should be smaller by several orders of magnitude.
\end{itemize}
In the remaining part of this paper, we explain further the details of possible experimental setup and calculate the transition rates. With the typical input parameters, the experimental sensitivity to the axion-photon coupling and that to the axion-electron coupling for a given axion mass are forecasted, as shown in Fig.~\ref{fig:sensitivity}.

\prlsection{Atomic Superradiance}{.}%
If the macroscopic coherence is established among target atoms or molecules, the rate of radiative emission will be enormously enhanced due to collective effects of all the atoms, i.e., the so-called superradiance (SR)~\cite{Dicke:1954zz,Gross:1982aa}. Compared to the intensity of photons from stochastic emission, the SR intensity will be proportional to the square of the radiant number $N^2$ rather than $N \sim \mathcal{O}(10^{23})$, due to the interference among different radiants. The idea of using SR to magnify the atomic transition rate was originally introduced to determine neutrino properties with the radiative emission of a neutrino pair~\cite{Yoshimura:2008ya, Yoshimura:2011ri, Fukumi:2012rn, Dinh:2012qb, Song:2015xaa, Zhang:2016lqp, Boyero:2015eqa, Vaquero:2016ovj, Huang:2019phr}. The emission rate of neutrino pairs can reach $\mathcal{O}(1)~{\rm s^{-1}}$ by using the trigger laser to irradiate the signal mode in a cm-scale atomic target, which should be very challenging for a realistic detection. By contrast, the stimulated capture or emission of cosmic axions may be more practicable due to the huge number density of axions when they constitutes all the dark matter with an energy density $\rho^{}_{\rm DM} \simeq 0.3~{\rm GeV\cdot cm^{-3}}$.

For the emission with a single outgoing photon, the coherence length is typically limited by its wavelength $\lambda$, and the final rate involves a diffraction factor  $(\lambda/L)^2$ with $L$ being the target length. Namely, only the radiants within the scale comparable to $\lambda$ can radiate collectively. Such a restriction can be relaxed if the initial prepared wave vector $\bm{k}^{}_{\rm eg}$ in the medium matches with that of the outgoing modes \cite{Scully:2006aa,Tanaka:2017juo}. For the paired superradiance (PSR)~\cite{Yoshimura:2008ya, Yoshimura:2012tm, Yoshimura:2014zha}, i.e. $|{\rm e}\rangle \to |{\rm g}\rangle + \gamma^{}_{1} + \gamma^{}_{2}$, the spatial phase factor of the two-photon emission amplitude is $\exp[{{\rm i} (\bm{k}^{}_{1}+\bm{k}^{ }_{2})\cdot \bm{x}^{}_i}]$ with $\bm{x}^{}_i$ being the coordinate vector of the $i$th radiant and $\bm{k}^{}_{1}$ ($\bm{k}^{ }_{2}$) being the wave vector of $\gamma^{}_{1}$ ($\gamma^{}_{2}$). By properly choosing the wave vectors such that $\bm{k}^{}_{1}+\bm{k}^{}_{2} = \bm{k}^{}_{\rm eg}$, the macroscopic coherence among all the target atoms can be guaranteed, but at the cost of a phase-space reduction that only the outgoing modes satisfying the momentum relation undergo PSR. The PSR is important to produce the topological soliton structure of electromagnetic waves~\cite{Yoshimura:2011ri, Yoshimura:2012tm, Yoshimura:2014zha} which can be utilized as the powerful stimulation of other processes and a possible way to suppress the electromagnetic background in the search of rare atomic transitions.

Very recently the PSR has been experimentally confirmed in Refs.~\cite{Miyamoto:2014aaa, Miyamoto:2015tva, Miyamoto:2017tva, Hiraki:2018jwu}, where the parahydrogen (${\rm p}$-${\rm H}^{}_{2}$) molecules have been used as the target and the explosive two-photon emission have been observed with a rate enhanced by a factor of $\mathcal{O}(10^{18})$ as compared to the stochastic spontaneous emission. The macroscopic coherence among the radiants in the target ensemble can be established by several approaches, e.g., the adiabatic Raman scattering~\cite{Miyamoto:2014aaa, Miyamoto:2015tva, Miyamoto:2017tva, Hiraki:2018jwu} and the technique of the coherent population return~\cite{Boyero:2015eqa, Vaquero:2016ovj}. The key point of the coherence preparation is to manipulate the atomic state in the superposition of the ground and excited states $\left| \psi \right> = \left(\left| \rm e \right>+\left| \rm g \right> \right)/ \sqrt{2}$. Hence the factor $\rho^{}_{\rm eg} \equiv \left< \rm e | \psi \right> \left< \psi | \rm g \right>$, whose maximum is $1/2$ by definition, is introduced to measure the coherence of the whole system.

\prlsection{Transition Amplitude}{.}%
As shown in Fig.~\ref{fig:transition1}, three interaction vertices are involved in the TREACA process $ |{\rm e}\rangle \to |{\rm g}\rangle + \gamma^{}_{1} + \gamma^{}_{2} + a$ or $a + |{\rm e}\rangle \to |{\rm g}\rangle + \gamma^{}_{1} + \gamma^{}_{2}$: (i) $\gamma$-$\bm d$ (or $\gamma$-$\bm \mu$), an external photon attached to one electric dipole $\bm d$ (or magnetic dipole $\bm \mu$); (ii) $a$-$\gamma$-$\gamma^{*}_{}$, the coupling of the axion with a virtual photon and the other external photon; (iii) $\gamma^*_{}$-$\bm d$ (or $\gamma^*$-$\bm \mu$), the virtual photon being attached to the other electric or magnetic dipole.
The vertices (ii) and (iii) are connected by the photon propagator associated with $\gamma^*$. The interaction vertex $a$-$\gamma$-$\gamma^{*}_{}$ is given by the first term in the Lagrangian in Eq.~(\ref{eq:Lagrangian}) and can be rewritten as the effective Hamiltonian~\cite{Raffelt:2006cw}
\begin{align}
\label{eq:Hamiltonianagg}
{\cal H}_{a \gamma\gamma}(x) = - g^{}_{a\gamma\gamma} a(x) {\bm{\mathcal{E}}(x)\cdot \bm{\mathcal{B}}}(x)  \;,
\end{align}
where $\bm{\mathcal{E}}(x)$ and $\bm{\mathcal{B}}(x)$ denote the electric and magnetic field, respectively. Hence the time-dependent Hamiltonian is just $H^{}_{a\gamma\gamma}(t)= \int  {\cal H}_{a \gamma\gamma}(t,{\bm x})\mathrm{d}^3 {\bm x} $. If the dark matter within our galaxy is solely composed of axions with a local energy density $\rho^{}_{\rm DM} \simeq 0.3~{\rm GeV\cdot cm^{-3}}$, the axions can be described by an oscillating classical field $a(x) = \sqrt{2\rho^{}_{\rm DM}}/m^{}_{a} \cdot \cos(\omega^{}_{a} t -  {\bm k^{}_{a}}{\cdot}{\bm x} )$, where $\omega^{}_{a} \simeq m^{}_{a}$ and ${\bm k^{}_{a}} \simeq m^{}_{a} {\bm v}^{}_{a}$ are the energy and momentum of the axion, respectively, with $|{\bm v}^{}_{a}| = v^{}_a \simeq \mathcal{O}(10^{-3})$ being the velocity. The interaction between the electromagnetic field and the atom is assumed to be mainly determined by the Hamiltonian ${H}_{\bm d}(t,\bm{x^{}_{d}}) =- {\bm d}\cdot \bm{\mathcal{E}}(t,\bm{x^{}_{d}})$ for the electric dipole or ${ H}_{\bm d}(t,\bm{x^{}_{d}}) =- {\bm \mu}\cdot \bm{\mathcal{B}}(t,\bm{x^{}_{\mu}})$ for the magnetic dipole, where $\bm{x^{}_{d}}$ or $\bm{x^{}_{\mu}}$ is the spatial location of the dipole in the target.

To be specific, the transitions $\left| \rm e \right> \rightarrow \left| \rm v \right> $ and $\left| \rm v \right> \rightarrow \left| \rm g \right> $ are both taken to be of the electric-dipole type, denoted as ${\bm d}^{}_{\rm ve}$ and  ${\bm d}^{}_{\rm vg}$, respectively. It is quite straightforward to extend the calculations to the case of magnetic-dipole transitions. Putting all together, we obtain the overall effective Hamiltonian
\vspace{-0.9cm}
\begin{widetext}
\begin{align}
\label{eq:Heffint}
-{  H}^{\rm int}_{\rm eff}(t,\bm{x^{}_{d}}) = &\left[ \frac{g^{}_{a \gamma \gamma} }{2m^{}_{a}} \sqrt{\frac{\rho^{}_{\rm DM}}{2}} {\bm d}^{}_{\rm ve} \cdot \bm{\mathcal{B}}(k)
\mathrm{e}^{-\mathrm{i}\left[(E^{}_{\rm ve} + \omega^{}_{} \pm \omega^{}_{a})t - ({\bm k}\pm {\bm k}^{}_{a}) \cdot \bm{x^{}_{d}} \right]}   + {\bm d}^{}_{\rm ve}\cdot \bm{\mathcal{E}}(k) \mathrm{e}^{-\mathrm{i} \left[ (E^{}_{\rm ve} + \omega^{}_{})t -  {\bm k}^{}_{}  \cdot \bm{x^{}_{d}} \right]} \right] \left| \rm v \right> \left< \rm e \right|  \\ \notag
& + \left[ \frac{g^{}_{a \gamma \gamma} }{2m^{}_{a}} \sqrt{\frac{\rho^{}_{\rm DM}}{2}}  {\bm d}^{}_{\rm vg} \cdot \bm{\mathcal{B}}(\tilde{k})\mathrm{e}^{-\mathrm{i}\left[(-E^{}_{\rm vg} + \tilde{\omega} \pm \omega^{}_{a})t -(  \tilde{{\bm k}} \pm {\bm k}^{}_{a}) \cdot \bm{x^{}_{d}}\right]} + {\bm d}^{}_{\rm vg}\cdot \bm{\mathcal{E}}(\tilde{k}) \mathrm{e}^{-\mathrm{i} \left[(-E^{}_{\rm vg} + \tilde{\omega})t -
	 \tilde{\bm k} \cdot \bm{x^{}_{d}}\right]
} \right] \left|\rm g \right> \left< \rm v \right| + {\rm h.c.}\;,
\end{align}
\end{widetext}
where $E^{}_{\rm ve}$ is the energy difference between $\left| \rm v \right>$ and $\left|\rm e \right>$ while $E^{}_{\rm vg}$ that between $\left| \rm v \right>$ and $\left| \rm g \right>$, and ``$+$" or ``$-$" in the phase factor refers to the emission or absorption of the axion. Moreover, $k =  (\omega, {\bm k})$ and $\tilde{k} = (\tilde{\omega}, \tilde{\bm k})$ stand for the four-momentum of the photon interacting with the electric dipole ${\bm d}^{}_{\rm ve}$ and ${\bm d}^{}_{\rm vg}$, respectively. The derivation of Eq.~(\ref{eq:Heffint}) with more details can be found in the supplemental material. Note that ${k} = k^{}_{1}$ and $\tilde{k} = k^{}_{2}$, or ${k} = k^{}_{2}$ and $\tilde{k} = k^{}_{1}$, should be satisfied, where $k^{}_1$ and $k^{}_2$ are the four-momenta of the photons $\gamma^{}_1$ and $\gamma^{}_2$. Some comments are helpful. First, if the Hamiltonian is integrated over time, the temporal phase will give rise to the condition of energy conservation. Second, the spatial phase is crucially important for the macroscopic coherence among multiple atoms, and the interference terms of different atomic dipoles should not be averaged out. Third, if one of the dipoles is of the magnetic type, we can simply replace $\bm{\mathcal{E}}$ (or $\bm{\mathcal{B}}$) by $\bm{\mathcal{B}}$ (or $-\bm{\mathcal{E}}$) for the electromagnetic field coupled to the magnetic dipole.

The final transition amplitude receives contributions from four possible contractions. Once one of two external photons $\gamma^{}_{1}$ is contracted with one of four electromagnetic field operators in Eq.~(\ref{eq:Heffint}), the contraction of the other photon $\gamma^{}_{2}$ will be uniquely determined. In the assumption of
$E^{}_{\rm eg} \ll E^{}_{\rm vg} \sim E^{}_{\rm ve}$, the transition amplitude at the leading order for the dipole at $\bm{x^{}_{d}}$ is
$\mathcal{M}(\bm{x^{}_{d}}) = \rho^{}_{\rm eg} \mathcal{M} \exp{\left[\mathrm{i}({\bm k}^{}_{1}+{\bm k}^{}_{2} \pm {\bm k}^{}_{a} -{\bm k}^{}_{\rm eg} ) \cdot \bm{x^{}_{d}} \right]} $ with
\begin{align}
\label{eq:amplitude1}
\mathcal{M} \simeq & \frac{g^{}_{a \gamma \gamma} \lambda^{}_{\theta}}{2m^{}_{a} E^{}_{\rm ve}} \sqrt{\frac{\rho^{}_{\rm DM}}{2}} \left|\bm{d}_{\rm vg}\right| \left| \bm{\mathcal{E}}^{}_{}(k^{}_{1}) \right|  \left|\bm{d}^{}_{\rm ve} \right| \left|\bm{\mathcal{E}}^{}_{}(k^{}_{2}) \right|  \;,
\end{align}
where $\lambda^{}_{\theta} \sim \mathcal{O}(1)$ is the polarization form factor that depends on polarizations of external photons and the dipoles. Here $\left|\bm{\mathcal{E}}(k^{}_{1})\right| = \sqrt{n^{}_{1} \omega^{}_{1}/2}$ and $\left|\bm{\mathcal{E}}(k^{}_{2})\right| = \sqrt{n^{}_{2} \omega^{}_{2}/2}$ are the electric-field strengths of the triggering lasers associated with $\gamma^{}_{1}$ and $\gamma^{}_{2}$, respectively, with $n^{}_{1}$ and $n^{}_{2}$ being the photon number densities. The dipole strength $\left|\bm{d}_{\rm vg}\right|$ is related to the transition rate $\gamma^{}_{\rm vg} = {E^{3}_{\rm vg} |\bm{d}^{}_{\rm vg}|^2}/({3\pi})$ for $\left|\rm v \right> \to \left|\rm g \right>$, where $\gamma^{}_{\rm vg}$ is just the Einstein A-coefficient for the spontaneous radiative emission. Therefore, we get
\begin{align}
\label{eq:amplitudesq}
|\mathcal{M}|^2 =
 \frac{9\pi^2}{32}
  \frac{g^{2}_{a\gamma\gamma} \rho^{}_{\rm DM}}{ m^2_a}
  \frac{\gamma^{}_{\rm vg}\gamma^{}_{\rm ve}n^{}_{\rm 1} n^{}_{\rm 2} \omega^{}_{\rm 1} \omega^{}_{\rm 2} \lambda^2_{\theta}}{E^3_{\rm vg}E^5_{\rm ve}} \; ,
\end{align}
where the transition rate $\gamma^{}_{\rm ve}$ and the energy difference $E^{}_{\rm ve}$ for $\left|\rm e \right> \to \left|\rm v \right>$ are defined similarly. The values of these parameters can only be fixed after the specific atomic system and the energy levels are chosen. The relevant information about the energy levels of different atoms can be found in Ref.~\cite{NIST}.

\prlsection{Transition Rate}{.}%
Then we proceed to estimate the stimulated emission and absorption rates in the atomic system with the SR effect. After summing over all the atoms in the target with a number density of $n^{}_{\rm tar}$,
i.e. $\mathcal{M}^{}_{\rm tot} = \sum^{}_{\bm{x^{}_{d}}} \mathcal{M}(\bm{x^{}_{d}})$, we obtain the total amplitude
\begin{align}
	\label{eq:amplitude2}
	\mathcal{M}^{}_{\rm tot} \simeq \rho^{}_{\rm eg} \mathcal{M} n^{}_{\rm tar} (2\pi)^3 \delta^3({\bm k}^{}_{1}+{\bm k}^{}_{2} \pm {\bm k}^{}_{a} -{\bm k}^{}_{\rm eg} )  \;,
\end{align}
where the momentum conservation ${\bm k}^{}_{1}+{\bm k}^{}_{2}\pm {\bm k}^{}_{a} -{\bm k}^{}_{\rm eg} = 0$ is guaranteed by the delta function that arises from the spatial integration over the target volume. It should be noticed that the maximal value of the delta function is now given by the target volume. When the momentum conservation is not fulfilled, e.g. ${\bm k}^{}_{1}+{\bm k}^{}_{2}\pm {\bm k}^{}_{a} -{\bm k}^{}_{\rm eg} = \Delta {\bm k}$, the phases of the transition amplitudes from different radiants are not matched, leading to the diffraction suppression $1/( L |\Delta {\bm k}|)^2$ in the final rate. For a realistic experimental setup, the frequencies of two lasers should be adjusted gradually to scan over the range determined by the unknown axion mass. During the scanning procedure, the axion absorption or emission rate will reach the peak value due to the macroscopic coherence when the momentum conservation is satisfied and the delta function is then simply replaced by the target volume $V^{}_{\rm tar}$.

The total rate for the whole ensemble reads $\Gamma^{}_{\rm tot} = |\mathcal{M}^{}_{\rm tot}|^2 2\pi \delta(\omega^{}_{1} + \omega^{}_{2} \pm \omega^{}_{a} - E^{}_{\rm eg})$, where the delta function for the energy conservation stems from the integration over the interaction time. For a given axion mass, the energy conservation can be achieved by tuning the laser frequencies. In the assumption of ${\bm k}^{}_{\rm eg} = 0$, the requirements for the momentum and energy conservations can be met if ${\bm k}^{}_{1} \simeq - {\bm k}^{}_{2}$ and $\omega^{}_{1}\simeq \omega^{}_{2}=(m^{}_{a}+E^{}_{\rm eg})/2$ hold in the case of the axion absorption, where the correction on the order of tiny axion momenta is neglected. The total rate turns out to be
\begin{align}\label{eq:totrate}
\Gamma^{}_{\rm tot}  =  \frac{9\pi^2}{32}
\frac{g^{2}_{a\gamma\gamma} \rho^{}_{\rm DM}}{ m^2_a}
\frac{\gamma^{}_{\rm vg}\gamma^{}_{\rm ve}n^{}_{\rm 1} n^{}_{\rm 2} \omega^{}_{\rm 1} \omega^{}_{\rm 2} \lambda^2_{\theta}}{E^3_{\rm vg}E^5_{\rm ve}} |\rho^{}_{\rm eg}|^2 N^2_{\rm tar} t^{}_{\rm min}
\;,
\end{align}
where ${g^{}_{a\gamma\gamma} }/{ m^{}_a}$ is a constant for a given axion model ($0.377~{\rm GeV}^{-2}$ for the KSVZ model and $0.139~{\rm GeV}^{-2}$ for the DFSZ model), $N^{}_{\rm tar} \equiv V^{}_{\rm tar} n^{}_{\rm tar}$ is the number of target atoms and $t^{}_{\rm min}$ is the minimum of several characteristic time scales: (i) the time duration of the triggering laser beam, which is inversely proportional to the full width at half maximum (FWHM) of the frequency distribution of the laser, typically on the order of $\mathcal{O}(\rm ns)$; (ii) the relaxation time of coherence in the medium; (iii) the inverse of the damping term for the transition; (iv) the coherence time of the axion field, i.e., $2\pi /(m^{}_{a}v^{2}_{a}) \simeq 413~{\rm ns}~(10^{-2}~{\rm eV}/m^{}_{a})$.
\begin{figure}[t]
	\begin{center}
		\includegraphics[width=0.36\textwidth]{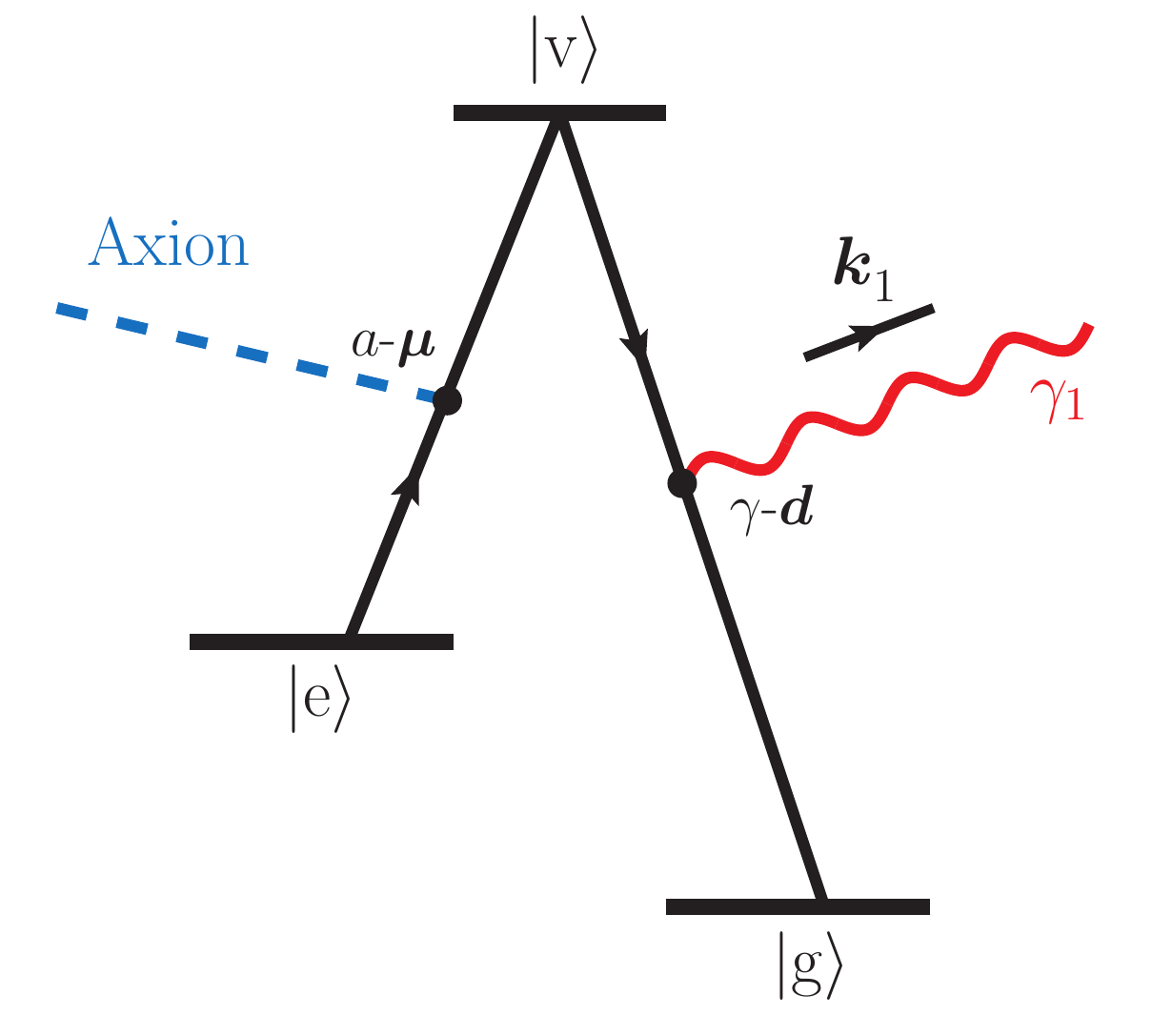}
	\end{center}
	\caption{The atomic transition $\left|\rm e\right>  \rightarrow \left| \rm g \right> + \gamma^{}_{1} + a$ or $\left|\rm e\right>+a \rightarrow \left| \rm g \right> + \gamma^{}_{1}$ induced by the axion-electron coupling.}
	\label{fig:transition2}
\end{figure}

Take the solid ${\rm p}$-${\rm H}^{}_{2}$ for example, for which $\left|\rm g \right>$ and $\left|\rm e \right>$ can be identified as its ground state and first vibrationally-excited state, as demonstrated in Ref.~\cite{Yoshimura:2017ghb,Yoshimura:2012tm}. For a nominal experimental setup with $N^{}_{\rm tar} =  10^{23}$, $t^{}_{\rm min} = 1~{\rm ns}$, $|\rho^{}_{\rm eg}| = 1/2$ (i.e., a complete coherence), $\omega^{}_{1} \simeq \omega^{}_{2} \simeq 0.26~{\rm eV}$, $n^{}_{1} \simeq n^{}_{2} \simeq 10^{18}~{\rm cm^{-3}}$, $E^{}_{\rm vg}\simeq E^{}_{\rm ve} \simeq 11~{\rm eV}$ and $\gamma^{}_{\rm vg}\gamma^{}_{\rm ve} \simeq 2.86 \times 10^{-4}~{\rm ns^{-2}}$ \cite{Yoshimura:2017ghb}, we obtain
\begin{align} \label{eq:rate1}
\Gamma^{\rm KSVZ}_{\rm tot}  &\simeq 9.62 \times 10^5~{\rm s^{-1}}, \nonumber \\
\Gamma^{\rm DFSZ}_{\rm tot}  &\simeq 1.30 \times 10^6~{\rm s^{-1}}
\;,
\end{align}
for the KSVZ and DFSZ models. With a relaxation time of coherence of $\mathcal{O}(10~{\rm ns})$, the total number of events for each deexcitation circle is $ 10^{-2}$ for the solid ${\rm p}$-${\rm H}^{}_{2}$ molecules, implying that at least $\mathcal{O}(100)$ deexcitation circles are required for $\mathcal{O}(1)$ signal photon emission. The number of trigger photons stored in the medium of one deexcitation circle is around $10^{18}$, which would be a severe intrinsic background for the signal detection. If the soliton condensate structure in the medium can be formed, the trigger background will be significantly reduced~\cite{Fukumi:2012rn}. According to the example in Ref.~\cite{Fukumi:2012rn}, a reduction factor about $10^{-8}$ can be achieved, but it is still far from enough for detecting the signal in ${\rm p}$-${\rm H}^{}_{2}$. The trigger laser photons can be trapped in the central part of the target, and only an exponentially-suppressed fraction of them leaks from the target ends. Therefore, one has to either invent a more powerful method of background reduction (as for ${\rm p}$-${\rm H}^{}_{2}$ a nearly background-free environment is required), or search for other atomic or molecular candidates with even larger signal rates.

\prlsection{Axion-electron Coupling}{.}%
The TREACA process can also take place via the direct coupling of axions with electrons~\cite{Yoshimura:2017ghb, Sikivie:2014lha, Borghesani:2015bla, Braggio:2017oyt}, as indicated by the second term of the Lagrangian in Eq.~(\ref{eq:Lagrangian}). Given the axion-electron coupling, the transition $\left|\rm e\right>  \rightarrow \left| \rm g \right> + \gamma^{}_{1} + a$ or $\left|\rm e\right>+a \rightarrow \left| \rm g \right> + \gamma^{}_{1}$ can occur similarly in the $\Lambda$-type system, as sketched in Fig.~\ref{fig:transition2}. First, the metastable level $\left|\rm e \right>$ jumps to a virtual state $\left|\rm v \right>$ by emitting or absorbing an axion through its derivative axial-vector coupling. Since the parity is conserved by this type of interaction, the ${\rm E1}$-type transition is forbidden and the dipole that connects $\left|\rm e \right>$ to $\left|\rm v \right>$ should be of the ${\rm M1}$-type. Second, the intermediate level $\left|\rm v \right>$ deexcites to the ground state $\left|\rm g\right>$ by emitting a single photon, which will be observed as the signal. The ${\rm E1}$ dipole transition should be adopted for $\left|\rm v \right> \rightarrow \left|\rm g \right>$ to enhance the total rate. The candidates like $\rm Yb$ or $\rm Xe$ are able to provide the required ${\rm M1} {\times} {\rm E1}$ transitions~\cite{Song:2015xaa}.
As an example, the following atomic levels of $\rm Yb$ will be selected: $\left|\rm g \right> = 4 f^{14}(^{1}\!S)6 s^2~^1\!S^{}_{J=0}$, $\left|\rm e \right>= 4 f^{14}(^{1}\!S)6 s 6p~^3\!P^{o}_{J=0}$ and $\left|\rm v \right> = 4 f^{14}(^{1}\!S)6 s 6p~^3\!P^{o}_{J=1}$.
The effective Hamiltonian for the interaction between the axion and the atomic electron can be written as
\begin{align}
\label{eq:Hamiltonianaspin}
{H}_{a \overline{e} e} ({\bm{x^{}_{s}}})=  - \frac{1}{2} g^{}_{a \overline{e} e} \bm{\nabla} a({\bm{x^{}_{S}}}) \cdot \bm{S} \;,
\end{align}
where $\bm{S}$ is the electron spin operator and ${\bm{x^{}_{S}}}$ is its spatial coordinate. Similarly one can find the transition amplitude $\mathcal{M}(\bm{x^{}_{S}}) = \rho^{}_{\rm eg} \mathcal{M}^{\prime} \exp{\left[\mathrm{i}({\bm k}^{}_{1} \pm {\bm k}^{}_{a} -{\bm k}^{}_{\rm eg} ) \cdot \bm{x^{}_{S}} \right]} $ with
 \begin{align}\label{eq:Mprime}
 {\cal M}^{\prime} = \frac{g^{}_{a \overline{e} e}}{2m_a E_{\rm ve} } \sqrt{\frac{\rho^{}_{\rm DM}}{2}}  (\bm{k}_a \cdot \bm{S}^{}_{\rm ve}) (\bm{d}_{\rm vg} \cdot \bm{\mathcal{E^{}_{\rm 1}}} ) \;,
 \end{align}
where $\bm{\mathcal{E^{}_{\rm 1}}}$ is the electric-field strength of the triggering laser photon and $\bm{S}^{}_{\rm ve}\equiv \left<\rm v \right|\bm{S}\left|\rm e \right>$ is the spin-flipped amplitude for $\left|\rm e\right> \to \left|\rm v\right>$. The axion mass $m^{}_a$ in Eq.~(\ref{eq:Mprime}) will be cancelled out by that in the momentum $\bm{k}^{}_a = m^{}_a \bm{v}^{}_a$ such that the final transition amplitude is independent of $m^{}_a$. With the number density $n^{}_{1}$ of laser photons, the transition rate can be calculated in the similar way as in Eq.~(\ref{eq:totrate}), namely,
\begin{align}
\label{eq:rateaee}
\Gamma^{\prime}_{\rm tot}
= \frac{3\pi  }{16}
\frac{g^{2}_{a \overline{e} e} \rho^{}_{\rm DM} \gamma^{}_{\rm vg} n^{}_{1} \omega^{}_{1} \lambda^2_{\theta} }{E^2_{\rm ve} E^3_{\rm vg}} |\bm{v}_a \cdot \bm{S}^{}_{\rm ve}|^2
 |\rho^{}_{\rm eg}|^2 N^2_{\rm tar} t^{}_{\rm min}
 \;,
\end{align}
where $|\bm{v}_a| \simeq \mathcal{O}(10^{-3})$ is the axion velocity.
For the $\rm Yb$ atom, the relevant parameters are $N^{}_{\rm tar} = 10^{23}$, $t^{}_{\rm min} = 1~{\rm ns}$, $|\rho^{}_{\rm eg}| = 1/2$, $\omega^{}_{1} \simeq 2.14~{\rm eV}$, $n^{}_{1} \simeq 10^{18}~{\rm cm^{-3}}$, $E^{}_{\rm vg}\simeq 2.23~{\rm eV}, E^{}_{\rm ve} \simeq 0.09~{\rm eV}$ and $\gamma^{}_{\rm vg} \simeq 1.15 \times 10^{-3}~{\rm ns^{-1}}$. The total rate in Eq.~(\ref{eq:rateaee}) is found to be
$10^7~{\rm s^{-1}}$ for the coupling constant $g^{}_{a \overline{e} e} = 10^{-13}~{\rm GeV^{-1}}$.

Since the macroscopic coherence demands the relation ${\bm k}^{}_{1} \pm {\bm k}^{}_{a} -{\bm k}^{}_{\rm eg} = 0$, the initial wave vector of the medium should be ${\bm k}^{}_{\rm eg} = {\bm k}^{}_{1} \pm {\bm k}^{}_{a} \simeq  {\bm k}^{}_{1}$, while $E^{}_{\rm eg} \simeq \omega^{}_{1} \pm m^{}_{a}$ holds for axion emission or absorption. If the two-photon absorption process is adopted to prepare the coherence~\cite{Tanaka:2017juo}, then $|{\bm k}^{}_{\rm eg}| \leq E^{}_{\rm eg}$ is required and thus only the axion emission with $E^{}_{\rm eg} \simeq \omega^{}_{1} + m^{}_{a}$ can be macroscopically enhanced. To form the soliton condensate of $\gamma^{}_{1}$ with PSR, another irradiating laser should be implemented. However, the previously chosen wave vector of the medium ${\bm k}^{}_{\rm eg} \simeq  {\bm k}^{}_{1}$ violates the new condition of momentum conservation when the additional laser photon is taken into account. As a consequence, a different set of atomic levels have to be selected to realize the soliton structure in order to reduce the intrinsic electromagnetic background.

\begin{figure}[t]
	\begin{center}
	\subfigure{
		\hspace{-0cm}
		\includegraphics[width=0.48\textwidth]{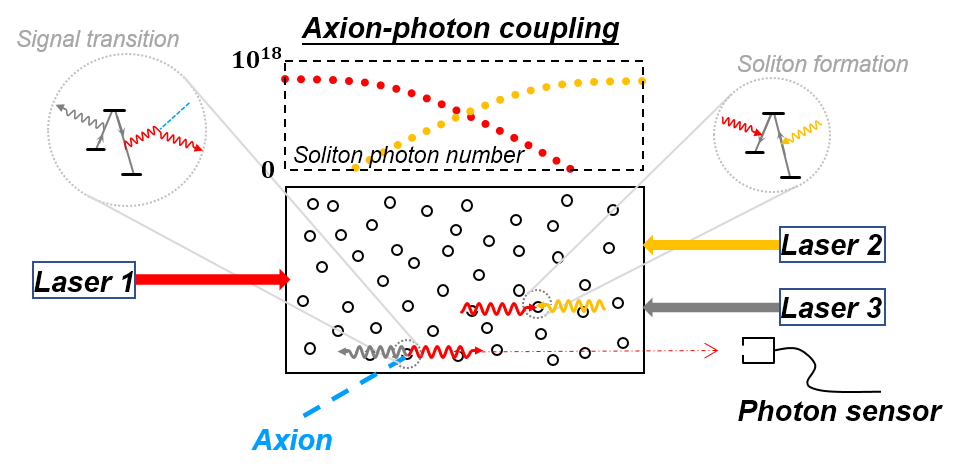}}
	\subfigure{
		\includegraphics[width=0.48\textwidth]{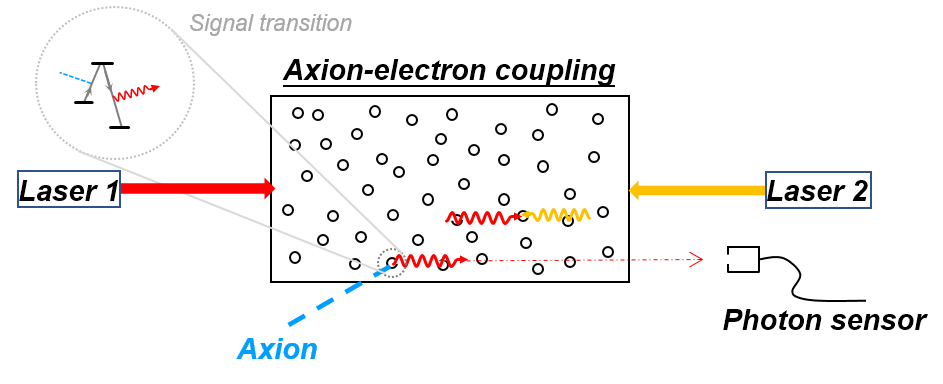}}
\end{center}
	\caption{The schematic diagram of the experimental setups to probe the axion-photon coupling (\emph{Upper Panel}) and axion-electron coupling (\emph{Lower Panel}). The lasers represented by the red and yellow colors are deployed to manufacture the soliton condensate. The transition by emitting or absorbing an axion can be identified by detecting the emitted photon that can escape from the soliton structure.}
	\label{fig:schematicDiagram}
\end{figure}

\prlsection{Conclusions}{.}%
In the ideal case of no background, we compute the $3\sigma$ sensitivities of such experiments to the axion-photon coupling $g^{}_{a\gamma\gamma}$ and the axion-electron coupling $g^{}_{a \overline{e} e}$ for a given axion mass range. A schematic diagram has been given in Fig.~\ref{fig:schematicDiagram}. The $3\sigma$ upper limits on the couplings have been obtained by requiring the number of signal events to be greater than 5.9, which is chosen as the average value for the Poisson probability of zero event to be $0.274\%$. The numerical results have been summarized in Fig.~\ref{fig:sensitivity} for the exposure time $T^{}_{\rm eff} = 10^{-6}~{\rm s}$, $10^{-4}~{\rm s}$ and $10^{-2}~{\rm s}$. For comparison, the sensitivities of several projected experiments have also been given. With an exposure of $10^{-2}~{\rm s}$, the TREACA experiment with an ideal setup can cover almost the entire parameter space of typical QCD axion models within the mass range of $(1 \cdots 10^4)~{\rm \mu eV}$, for which axions can constitute the whole cold dark matter.

Finally, we give further comments on the background, which remains an unsolved issue. In the present setup, the stimulated emission of photon signals will be the same as photons from the trigger lasers. Therefore, the soliton structure, within which the photons from PSR are effectively trapped in the medium, must be formed to efficiently suppress this intrinsic background at the target edge. If the target is long enough in size, the background may be negligible. However, this cannot be achieved with the two photons and the atomic levels in the TREACA process, simply due to that the energy and momentum relations of TREACA and PSR cannot be simultaneously satisfied. Another laser $\gamma^{}_{3}$ with a proper frequency may be required to establish the soliton structure. One may also consider the waveguide to remove the higher-order electromagnetic background (e.g., $3\gamma$ emission~\cite{Tanaka:2016wir}). To go beyond the perturbative calculations presented in this paper, we must solve the coupled Maxwell-Bloch equations in a sophisticated way, which may be left for future works. The approaches in this work can be applied also to experimental searches for other types of light dark matter coupled to the atomic or molecular system~\cite{Arvanitaki:2017nhi, Yang:2016zaz,Stadnik:2018sas}.

\prlsection{Acknowledgement}{.}%
{\sl We would like to thank Prof.~Motohiko Yoshimura for valuable communications. This work was supported in part by the National Natural Science Foundation of China under grant No.~11775232 and No.~11835013, and by the CAS Center for Excellence in Particle Physics.
}

\begin{appendix}
\section{Appendix}	
In this Appendix, we present some details about the derivation of the effective Hamiltonian and the calculations of the transition amplitudes.

\prlsection{Propagator Calculation}{.}%
As indicated in Fig.~1 of the paper, the vertices $\gamma^*_{}$-$\bm d$ and $a$-$\gamma$-$\gamma^{*}_{}$ are connected by the photon propagator. We take the time-ordered product of these two vertices, and add up all possible Wick contractions
\begin{align}\label{eq:timeordering}
&  {\cal T} \{ {\bm{\mathcal{E}}(x)\cdot \bm{\mathcal{B}}}(x) ~ {\bm{\mathcal{E}}(y)}
\cdot \bm{d}(y)  \}   = \\ \notag
&   \contraction{ }{ \bm{\mathcal{E}} } {(x)\cdot \bm{\mathcal{B}}(x)~ }  { \bm{\mathcal{E}}}
{\bm{\mathcal{E}}(x)\cdot \bm{\mathcal{B}}(x)} ~ {\bm{\mathcal{E}}(y)}
\cdot \bm{d}(y)
+
\contraction{ \bm{\mathcal{E}}(x) \cdot} { \bm{\mathcal{B}} ~}{(x)}{ \bm{\mathcal{E}}}
\bm{\mathcal{E}}(x)\cdot \bm{\mathcal{B}}(x) ~ \bm{\mathcal{E}}(y)
\cdot \bm{d}(y) \;,
\end{align}
where the electric-dipole coupling is assumed. By using the following expressions of propagators
\begin{align}\label{eq:ee}
& \left< 0 \right| {\cal T} \{ \bm{\mathcal{E}}^{i}_{}(x) {\bm{\mathcal{E}}}^{j}_{}(y)  \}  \left| 0 \right>
=  \\ \notag
& \hspace{1cm} \int \frac{\mathrm{d}^4 q}{(2\pi)^4} \mathrm{e}^{-{\rm i}q \cdot(x-y)} \frac{-{\rm i}(\bm{q}^i\bm{q}^j \cdot \omega^2/|\bm{q}|^2-\omega^2 \delta^{ij})}{\omega^2-|\bm{q}|^2}\;, \\
& \left< 0 \right| {\cal T} \{ \bm{\mathcal{E}}^{i}_{}(x) {\bm{\mathcal{B}}}^{j}_{}(y)  \}  \left| 0 \right>
=  \\ \notag
& \hspace{1cm} \int \frac{\mathrm{d}^4 q}{(2\pi)^4} \mathrm{e}^{-\mathrm{i}q \cdot(x-y)} \frac{-\mathrm{i}\epsilon^{}_{ijl}\omega \bm{q}^{l}_{}}{\omega^2-|\bm{q}|^2}\;, \\
& \left< 0 \right| {\cal T} \{ \bm{\mathcal{B}}^{i}_{}(x) {\bm{\mathcal{B}}}^{j}_{}(y)  \}  \left| 0 \right>
=  \\ \notag
& \hspace{1cm} \int \frac{\mathrm{d}^4 q}{(2\pi)^4} \mathrm{e}^{-\mathrm{i}q \cdot(x-y)} \frac{-\mathrm{i}(\bm{q}^i \bm{q}^j - |\bm{q}|^2 \delta^{ij})}{\omega^2-|\bm{q}|^2},
\end{align}
we can obtain
\begin{align}\label{eq:timeordering1}
&  {\cal T} \{ {\bm{\mathcal{E}}(x)\cdot \bm{\mathcal{B}}}(x) ~ {\bm{\mathcal{E}}(y)}
\cdot \bm{d}(y)  \}
\simeq \\ \notag
& \hspace{1cm} \mathrm{i} \int \frac{\mathrm{d}^4 q}{(2\pi)^4} \mathrm{e}^{-\mathrm{i}q \cdot(x-y)} \frac{\bm{\mathcal{B}}(x) \cdot \bm{d}(y)}{2} \;,
\end{align}
where $\epsilon^{}_{ijl}$ denotes the Levi-Civita tensor with the convention $\epsilon^{}_{123}=-1$, $q = (q^{}_{0}, \bm{q})$ is the four-momentum of the propagator. Note that the relation $q^{}_{0} \simeq |\bm{q}|+m^{}_{a}$, which holds as a very good approximation because of extremely non-relativistic axions, has been used to derive the above result. If the vertex is of the ${\rm M1}$ type $\gamma^*$-$\bm{\mu}$, one can simply replace $\bm{\mathcal{B}}(x)$ with $-\bm{\mathcal{E}}(x)$ in Eq.~(\ref{eq:timeordering1}). Note that our results in Eq.~(\ref{eq:timeordering1}) differ from those in Ref.~\cite{Yoshimura:2017ghb}, where it seems that only one Wick contraction in the time-ordered product in Eq.~(\ref{eq:timeordering}) has been considered.

\prlsection{Effective Hamiltonian}{.}%
The effective Hamiltonian for the atomic transition $\left|\rm e \right> \rightarrow \left|\rm v \right> + \gamma^{}_{} + a$ (similarly for $\left|\rm v \right> \rightarrow \left|\rm g \right> + \gamma^{}_{} + a$) can be determined by using the perturbation theory. Given the interaction Hamiltonians for the vertices $a$-$\gamma$-$\gamma^{*}_{}$ and $\gamma^*_{}$-$\bm d$, we can calculate the transition matrix ${\rm i}T \equiv S - 1$, with $S$ being the total scattering matrix, at the second order in the presence of the axion and laser background fields, namely,
\begin{align}\label{eq:smatrix}
{\rm i}T \simeq -\mathcal{T}\int_{-\infty}^{+\infty}\mathrm{d}t^{}_{a}
\int_{-\infty}^{+\infty}\mathrm{d}t^{}_{\bm d}
H^{}_{a\gamma\gamma}(t^{}_{a}) H^{}_{\bm d}(t^{}_{\bm d}),
\end{align}
where $t^{}_{a}$ (or $t^{}_{\bm d}$) is the time variable of the vertex $a$-$\gamma$-$\gamma^{*}_{}$ (or $\gamma^*_{}$-$\bm d$). Performing the integration over $t^{}_{a}$, we get
\begin{eqnarray}\label{eq:smatrix1}
\left<\rm v \right| {\rm i}T \left|\rm e \right> & \simeq  & \mathrm{i} \int_{-\infty}^{+\infty} \mathrm{d}t^{}_{\bm d}
\frac{g^{}_{a \gamma \gamma} }{2m^{}_{a}} \sqrt{\frac{\rho^{}_{\rm DM}}{2}} {\bm d}^{}_{\rm ve} \cdot \bm{\mathcal{B}}(k) \\ \notag
& &\times ~ \mathrm{e}^{-\mathrm{i}\left[(E^{}_{\rm ve} + \omega^{}_{} + \omega^{}_{a})t^{}_{\bm d} - ({\bm k}^{}_{}+ {\bm k}^{}_{a}) \cdot \bm{x^{}_{d}} \right]}\;
\end{eqnarray}
for $\left|\rm e \right> \rightarrow \left|\rm v \right> + \gamma^{}_{} + a$, where the following plane-wave expansions of the background axion and laser fields have been used
\begin{eqnarray}\label{eq:planewave}
a(x) & = & \sqrt{\frac{\rho^{}_{\rm DM}}{2}}\frac{1}{m^{}_{a}} \cdot
\left(
\mathrm{e}^{-\mathrm{i} k^{}_{a} \cdot x} +\mathrm{e}^{\mathrm{i} k^{}_{a} \cdot x} \right) \;, \\
\bm{\mathcal{B}}(x) & = & \bm{\mathcal{B}}(k) \left(
\mathrm{e}^{-\mathrm{i} k^{}_{} \cdot x} +\mathrm{e}^{\mathrm{i} k^{}_{} \cdot x} \right)\;.
\end{eqnarray}
To obtain the same transition amplitude in Eq.~(\ref{eq:smatrix1}), we can identify the effective Hamiltonian that induces the transition
$\left|\rm e \right> \rightarrow \left|\rm v \right> + \gamma^{}_{} + a$
as
\begin{eqnarray}\label{eq:Heff}
&-&
\frac{g^{}_{a \gamma \gamma} }{2m^{}_{a}} \sqrt{\frac{\rho^{}_{\rm DM}}{2}} {\bm d}^{}_{\rm ve} \cdot \bm{\mathcal{B}}(k)
\mathrm{e}^{-\mathrm{i}\left[(E^{}_{\rm ve} + \omega^{}_{} + \omega^{}_{a})t^{}_{\bm d} - ({\bm k}+ {\bm k}^{}_{a}) \cdot \bm{x^{}_{d}} \right]}\; \nonumber \\
& +& {\rm h.c.} \;
\end{eqnarray}
Similarly, one can obtain the effective Hamiltonian for $\left|\rm v \right> \rightarrow \left|\rm g \right> + \gamma^{}_{} + a$. These results are then adopted in Eq.~(3) of the paper.

\prlsection{Amplitude}{.}%
The transition amplitude for $|{\rm e}\rangle \to |{\rm g}\rangle + \gamma^{}_{1} + \gamma^{}_{2} + a$ can readily be figured out by implementing the effective interaction Hamiltonian in Eq.~(3) to the atomic system. More explicitly, the matrix element $\left<\rm g \right| {\rm i}T \left|\rm e \right> $ is given by
\begin{align}\label{eq:smatrixeg}
\mathcal{T} \frac{(-\mathrm{i})^2}{2!}\int_{-\infty}^{+\infty}\mathrm{d}t
\int_{-\infty}^{+\infty}\mathrm{d}\tilde{t}
\left<\rm g \right|{  H}^{\rm int}_{\rm eff}(t) {  H}^{\rm int}_{\rm eff}(\tilde{t}) \left|\rm e \right> \;.
\end{align}
Since the transition is accomplished via an intermediate state $\left|\rm v \right>$, the matrix element is then written as
\begin{align}\label{eq:smatrixegv}
(-\mathrm{i})^2\int_{-\infty}^{+\infty}\mathrm{d}\tilde{t}
\int_{-\infty}^{\tilde{t}}\mathrm{d}t
\left<\rm g \right|{  H}^{\rm int}_{\rm eff}(\tilde{t}) \left|\rm v \right>\left<\rm v \right| {  H}^{\rm int}_{\rm eff}(t) \left|\rm e \right> \;.
\end{align}
After carrying out the time integration in Eq.~(\ref{eq:smatrixegv}), one can extract the amplitude from $\left<\rm g \right| {\rm i}T \left|\rm e \right> = \mathrm{i} \mathcal{M} \times 2\pi \delta(E^{}_{\rm eg} - \omega^{}_{1}-\omega^{}_{2} - \omega^{}_{a})$ as below
\begin{eqnarray}
\label{eq:amplitude3}
\mathcal{M} & \simeq & \frac{g^{}_{a \gamma \gamma}}{2m^{}_{a} E^{}_{\rm ve}} \sqrt{\frac{\rho^{}_{\rm DM}}{2}}  \left[
\bm{d}^{}_{\rm vg} \cdot \bm{\mathcal{E}}^{}_{}(k^{}_{1})~\bm{d}^{}_{\rm ve} \cdot \bm{\mathcal{B}}^{}_{}(k^{}_{2})  \right. \nonumber \\
&~&  +\bm{d}^{}_{\rm vg} \cdot \bm{\mathcal{E}}^{}_{}(k^{}_{2})~\bm{d}^{}_{\rm ve} \cdot \bm{\mathcal{B}}^{}_{}(k^{}_{1}) \nonumber \\
&~& +  \bm{d}^{}_{\rm vg} \cdot \bm{\mathcal{B}}^{}_{}(k^{}_{1})~\bm{d}^{}_{\rm ve} \cdot \bm{\mathcal{E}}^{}_{}(k^{}_{2})  \nonumber \\
&~&  \left. + \bm{d}^{}_{\rm vg} \cdot \bm{\mathcal{B}}^{}_{}(k^{}_{2})~\bm{d}^{}_{\rm ve} \cdot \bm{\mathcal{E}}^{}_{}(k^{}_{1})
\right] \notag \\
& \equiv & \frac{g^{}_{a \gamma \gamma}\lambda^{}_{\theta}}{2m^{}_{a} E^{}_{\rm ve}} \sqrt{\frac{\rho^{}_{\rm DM}}{2}}
\left|\bm{d}_{\rm vg}\right| \left| \bm{\mathcal{E}}^{}_{}(k^{}_{1}) \right|  \left|\bm{d}^{}_{\rm ve} \right| \left|\bm{\mathcal{E}}^{}_{}(k^{}_{2}) \right|
\;, \quad
\end{eqnarray}
where the factor $1/E^{}_{\rm ve}$ can be interpreted as the lifetime of the virtual state $\left| \rm v \right>$ according to the time-energy uncertainty relation, and $E^{}_{\rm ve} \gg \omega^{}_{1},\omega^{}_{2}, m^{}_{a}$ is assumed. For ${\rm p}$-${\rm H}^{}_{2}$ with $\left|\rm g \right>$ and $\left|\rm e \right>$ being its ground state and first vibrationally-excited state, the form factor of polarization $\lambda^{}_{\theta}$ is proportional to $\sin{\theta^{}_{12}}$, where $\theta^{}_{12}$ is the relative angle between the electric-field polarizations of two triggering lasers~\cite{Yoshimura:2017ghb}.
\end{appendix}


\end{document}